\newcommand{\der}{ \partial \mskip -9mu / \mskip 4mu}
\newcommand{\Akr}{ A \mskip -9mu / \mskip 4mu}
\newcommand{\fkr}{ f \mskip -9mu / \mskip 4mu}
\newcommand{\pkr}{ p \mskip -9mu / \mskip 4mu}
\newcommand{\inkr}{ \in \mskip -14mu / \mskip 9mu}
\documentstyle[12pt]{article}

\title{Contour gauges, canonical formalism and flux algebras.}
\begin{document}
\begin{center}
\begin{Large}
{\bf Contour gauges, canonical formalism and flux algebras.}
\end{Large}
\end{center}
\vspace*{1cm}
\begin{center}
L.{\L}ukaszuk$^{a,}$\footnote{E-mail: lukaszuk@fuw.edu.pl},
E.Leader$^{b,}$\footnote{E-mail: e.leader@bbk.ac.uk},
A.Johansen$^{c,}$\footnote{E-mail: johansen$@$pauli.harvard.edu}
\end{center}
$^{a}$Soltan Institute for Nuclear Studies, Hoza 69, 00-681 Warsaw,
Poland \\
$^{b}$Birkbeck College, University of London, Malet Street, London
WC1E 7HX, England \\
$^{c}$The St.Petersburg Nuclear Physics Institute, Gatchina, 188350,
Russian Federation \\
\vspace{0.5cm} 
\begin{center}
We would like to dedicate this paper to the memory of our dear friend and
colleague Alyosha Anselm.
\end{center}
\begin{abstract}
A broad class of contour gauges is shown to be
determined by admissible contractions of the geometrical
region considered and a suitable equivalence class of curves
is defined. In the special case of magnetostatics, the relevant
electromagnetic potentials are directly related to the ponderomotive
forces. Schwinger's method of extracting a gauge invariant factor from the fermion propagator
could, it is argued, lead to incorrect results.
Dirac brackets of both Maxwell and Yang-Mills theories are given
for arbitrary admissible space-like paths. 
It is shown how to define a non-abelian flux and local charges which obey a local
charge algebra.
Fields associated with the charges differ from the electric fields of 
the theory by singular topological terms;
to avoid this obstruction to the Gauss law it is necessary to exclude a single,
gauge fixing curve from the region considered. 
\end{abstract}
\newpage
\section{Introduction}
The Fock-Schwinger gauge \cite{l1,l2}
\begin{equation}
A\cdot x = 0
\label{e1}
\end{equation}
has two remarkable features.\\
Firstly, the potentials can be expressed in terms of the fields
\cite{l3}:
\begin{equation}
A_{\mu}^{F-S}(x)= \int_{0}^{1} G_{\nu \mu}(x t) x^{\nu} t dt
\label{e2}
\end{equation}
where
\begin{equation}
G_{\mu \nu} = F_{\mu \nu} + i g [A_{\mu}, A_{\nu}]
\label{e'2}
\end{equation}
Equation (\ref{e2}) is true both in Maxwell and Yang-Mills theories
provided that the Bianchi identities are satisfied. \\
Secondly, the integral along {\it the straight line}
segment ${\em s}(x, 0)$ running from $0$ to $x$.
\begin{equation}
\varphi_{{\em s}(x, 0)} [A^{F-S}]= \int_{{\em s} (x, 0)} A_{\mu}
^{F-S}(y) dy^{\mu}
\label{e3}
\end{equation}
is constant i.e. independent of $x$
\begin{equation}
d\varphi_{{\em s}(x,0)}= \frac{\partial \varphi_{s}}{\partial x^{\mu}}
dx^{\mu} = 0
\label{e4}
\end{equation}
therefore
\begin{equation}
\varphi_{{\em s}(x, 0)}= \varphi_{{\em s}(0,0)} \equiv 0
\label{e4a}
\end{equation}
Eqn.(\ref{e4})  
can be also derived from (\ref{e3}) by direct calculation with the Bianchi
identities taken into account and will be used by us in a more general context
throughout this paper. \\
Equation (\ref{e4}) has an interesting physical interpretation. Take
electrodynamics and the null loop ${\em s}(x,0){\em \bar{s}}(x,0)$
i.e.
the path followed along the straight segment $(0,x)$ and back.
The flux $\Phi({\em s}_{x}\overline{{\em s}_{x}})= 0$ for any regular
field $F_{\mu\nu}$. Next consider the deformed loop consisting of
${\em s}(x+\Delta x, 0) {\em s}(x, x+\Delta x) {\em \bar{s}}(x, 0)$.
The change
of flux is gauge invariant and reads as:
\begin{equation}
\Delta\Phi = \varphi_{{\em s}(x+\Delta x, 0)}[A] - \varphi_{{\em s}
(x, 0)}[A] + \varphi_{{\em s}(x, x+\Delta x)}[A]
\label{e5}
\end{equation}
Now $\varphi_{{\em s}(0,x)}$ does not depend on x for $A = A^{F-S}$, so
\begin{equation}
\Delta\Phi = \varphi_{{\em s}(x, x+\Delta x)}[A^{F-S}]
\label{e5.1}
\end{equation}
Taking $\Delta x \longrightarrow 0$ we get the infinitesmal change of
the flux:
\begin{equation}
\Delta \Phi = - A_{\mu}^{F-S}(x) \Delta x^{\mu}
\label{e6}
\end{equation}
Therefore, $A_{\mu}^{F-S}(x)$ can be interpreted as
a generalization of the magnetostatic ponderomotive field. Indeed,
consider a closed circuit (current intensity I) which contains straight
segment ${\em s}(\vec{x}, 0)$. Deformation of ${\em s}(\vec{x}, 0)
\longrightarrow {\em s}(\vec{x}+\Delta \vec{x}, 0) {\em s}(\vec{x},
\vec{x}+
\Delta \vec{x})$ changes the magnetic interaction energy
by $-I  \Delta \Phi$ i.e. the ponderomotive force is $I A_{i}^{F-S}
(\vec{x})$ in this three-dimensional example.\\

The Fock-Schwinger gauge is a special case of contour gauges \cite{ll1},
which in our paper will be specified by condition (\ref{e4}) with $s(x, 0)$
replaced by more general family of curves $c(x, x_{0})$. Let us notice
that for any choice of $c(x, x_{0})$ the corresponding potential $f^{c}$
is again the ponderomotive field corresponding
to the deformation $c(x, x_{0}) \bar{c}(x, x_{0}) \longrightarrow
c(x+\Delta x, x_{0}) {\em s}(x, x+\Delta x) \bar{c}(x, x_{0})$.
Therefore it seems natural to name these path dependent gauges as
ponderomotive ones.
The ponderomotive interpretation nicely fits the fact that the contour
gauges are a part of larger family of physical gauges; in electrodynamics
a representant $A^{D}$ of physical gauge we characterise by choice of the
projection operator $\raisebox{2.0mm}{{\tiny D}}\hat{P}$
\begin{equation}
\raisebox{2.0mm}{{\tiny D}}\hat{P}^{\nu}_{\mu}
\cdot \raisebox{2.0mm}{{\tiny D}}\hat{P}^{\rho}_{\nu} =
\raisebox{2.0mm}{{\tiny D}}\hat{P}^{\rho}_{\mu}
\label{dzi1}
\end{equation}
satisfying
\begin{equation}
\raisebox{2.0mm}{{\tiny D}}\hat{P}^{\nu}_{\mu} \cdot (\partial_{\nu} \chi) = 0
\label{dzi2}
\end{equation}
for the arbitrary scalar function $\chi(x)$. \\
Then
\begin{equation}
{\cal A}^{(D)}_{\mu} \equiv \raisebox{2.0mm}{{\tiny D}}\hat{P}^{\nu}_{\mu} {\cal A}_{\nu}
\label{dzi3}
\end{equation}
In particular
\begin{equation}
{\cal A}^{(D)}_{\mu} \equiv \raisebox{2.0mm}{{\tiny D}}\hat{P}^{\nu}_{\mu}
{\cal A}^{(D)}_{\nu}
\label{dzi4}
\end{equation}
It follows from eqns.(\ref{dzi1}-\ref{dzi3}) that ${\cal A}^{(D)}$ is gauge invariant
quantity with respect to transformation of the gauge field $A_{\mu}(x)$, but, 
on the other hand the eqn. (\ref{dzi4}) is a gauge condition
for the choice of ${\cal A}^{(D)}$. A vast choice of different gauge invariant
formalisms \cite{NX} has evidently it's source in the freedom of defining
projection operators satisfying eqn.(\ref{dzi2}). \\
In this paper we shall present results concerning contour gauges,
characterized by a broad class of curves $c(x, x_{0})$; the corresponding
projection operator $P^{c(x,x_0)}$ acts on the 4-potentials as follows
\begin{equation}
[\hat{P}^{c} \cdot {\cal A}]_{\mu}(x) = {\cal A}_{\mu}(x) -
\frac{\partial}{\partial x^{\mu}} \int_{Y \in c(x,x_{0})} dY^{\nu}
{\cal A}_{\nu}(Y)
\label{dzi5}
\end{equation}
and as a result of eqn.(\ref{dzi4}) the chosen potential $f^{c(x,x_0)}$
has to satisfy eqn.(\ref{e4}) (with $s(x,0)$ replaced by $c(x,x_0)$).
In chapter 2 we show that for quite general class of curves (including
non-smooth ones) our gauge condition determines $f^{c}[F]$ in the case of
electrodynamics. In the same chapter a connection between possible contractions
of the space-time region considered and the freedom of choice of our
ponderomotive potentials is exhibited and a suitable equivalence class of
curves is defined.
A subclass of curves - suitable for Yang-Mills
theory - is specified and applied in Ch.3.
Some gauges which have been considered in this theory, such as
Fock-Schwinger \cite{l3}, superaxial \cite{l4}, temporal with
space-like Fock-Schwinger \cite{l5} are shown to
belong to our class of gauges.\\

We generalize - in Ch.5 - Schwinger's method \cite{l2} of extracting
a field dependent factor from the fermion propagator for a Dirac
particle interacting with a given electromagnetic field. This enables
us to provide an expansion for the fermion propagator in terms of the
electromagnetic fields themselves.\\
The consistency of the ponderomotive gauge constraints
with the canonical
formalism is exhibited in the Maxwell theory in Ch.6. The Dirac
brackets can be obtained in a straightforward way due to the fact
that our gauge constraint at fixed time is canonically conjugate
to the Gauss law constraint.\\

The application of canonical formalism to Yang-Mills theories is discussed in
chapters 7 - 9. The Dirac brackets are given in Ch.7.
Flux operator algebras and local charge algebras
with structure constants of underlying Yang-Mills theory are derived
in chapters 8, 9. Fields associated with these charges differ from 
the electric fields of the theory by singular topological terms;
to avoid this obstruction to the Gauss law it is necessary to exclude
a single gauge fixing curve from the region considered in the theory.

\section{Ponderomotive gauges - electrodynamics}
\setcounter{equation}{0}
Let us consider an arbitrary electromagnetic field
\begin{equation}
F_{\mu\nu}(x)=\partial_{\mu}A_{\nu}-\partial_{\nu}A_{\mu}
\label{e7}
\end{equation}
which is continuously once differentiable
at any $x\in V_{4}$, $V_{4}$ being
some open, simply connected space-time region. Next, let us associate
with every pair $(x_{1}, x_{2})$ a \underline{unique}
continuous path running from
$x_{2}\in V_{4}$ to $x_{1}\in V_{4}$ and contained in $V_{4}$:
\begin{equation}
(x_{1}, x_{2}) \longrightarrow c(x_{1},x_{2}) \subset V_{4}
\label{e8}
\end{equation}
The parametric representation of c is taken to be:
\begin{equation}
x\in c(x_{1},x_{2}) \Longleftrightarrow x^{\mu}=c^{\mu}(x_{1},x_{2},
\tau) \in V_{4}
\label{e9}
\end{equation}
\hspace{2cm} for some $\tau \in [0,1]$
with
\begin{eqnarray}
x_{2}^{\mu}=c^{\mu}(x_{1},x_{2},0) \nonumber \\
x_{1}^{\mu}=c^{\mu}(x_{1},x_{2},1)
\label{e10}
\end{eqnarray}
Our results will not depend on the equivalent reparametrizations:
\begin{equation}
\tau \longrightarrow t(\tau)
\end{equation}
with
\begin{equation}
\frac{\partial t}{\partial \tau} > 0,
\label{e11}
\end{equation}
In what follows we shall fix $x_{2}=x_{0}$ and demand that $c^{\mu}
(x,x_{0},\tau)$ satisfy the following regularity conditions:
\begin{itemize}
\item[i)] $c^{\mu}$ is $C^{(2)}$ - regular as a function of $x, \tau$
everywhere in $V_{4}\times[0,1]$ with the possible exception of $\tau=
\tau_{1},\ldots \tau_{n}$; $\tau_{i}(x)$ being $C^{(2)}$-regular.
\item[ii)] At $\tau \longrightarrow \tau_{i}$, at fixed x, the left -
and right - side finite limits exist for $n \leq 2$ derivatives.
These limits coincide for:
\begin{equation}
c^{\mu}(x,x_{0},\tau)\vert_{\tau \rightarrow \tau_{i}(x)}=c^{\mu}
(x,x_{0},\tau_{i}(x))
\label{e12}
\end{equation}
\begin{equation}
\frac{\partial c^{\mu}(x,x_{0},\tau)}{\partial x^{\rho}}\vert_
{\tau \rightarrow \tau_{i}}+\frac{\partial c^{\mu}}{\partial {\tau}}
\vert_{\tau \rightarrow \tau_{i}} \frac{\partial \tau_{i}(x)}
{\partial x_{\rho}} =
\frac{\partial c^{\mu}(x,x_{0},\tau_{i}(x))}{\partial x^{\rho}}
\label{e13}
\end{equation}
\item[iii)]In the limit $\tau \longrightarrow 0, 1$
\begin{equation}
c^{\mu}(x,x_{0},\tau ) \stackrel{\tau \rightarrow 1}{\longrightarrow}
x^{\mu}+O(1-\tau) g^{\mu}_{1}(x)
\label{e14}
\end{equation}
\begin{equation}
c^{\mu}(x,x_{0},\tau) \stackrel{\tau \rightarrow 0}{\longrightarrow}
x^{\mu}_{0}+O(\tau)g^{\mu}_{2}(x)
\label{e15}
\end{equation}
\end{itemize}
The somewhat general condition ii) is of importance; it admits
functions with discontinuously changing direction. \\

We shall now prove the following theorem:
\newtheorem{th1}{Theorem}
\begin{th1}
Given the electromagnetic field $F_{\mu\nu}$, the condition
\begin{equation}
d\varphi_{c(x,x_{0})}=0
\label{e16}
\end{equation}
with
\begin{equation}
\varphi_{c(x,x_0)}=\int_{c(x,x_{0})} A_{\mu}(y) dy^{\mu}
\label{e17}
\end{equation}
uniquely determines the electromagnetic potential A as
\begin{equation}
A_{\rho} = f_{\rho}^{c}(x) \equiv \int_{c(x,x_{0})} F_{\nu\mu}(y)
\frac{\partial y^{\mu}}{\partial x^{\rho}} dy^{\nu}
\label{e18}
\end{equation}
\end{th1}
{\it \underline{Proof}}\\
We can write
\begin{equation}
\varphi_{c(x,x_{0})}= \sum_{i=0}^{n} I_{i}
\end{equation}
with
\begin{equation}
I_{i}=\int_{\tau_{i}}^{\tau{i+1}} A_{\mu}(c) \frac{\partial c^{\mu}}
{\partial \tau} d\tau
\end{equation}
and
\begin{eqnarray}
\tau_{0}=0 \nonumber \\
\tau_{n+1}=1
\end{eqnarray}
Next,
\begin{equation}
\frac{\partial I_{i}}{\partial x^{\rho}}= A_{\mu}(c) \frac
{\partial c^{\mu}}{\partial \tau} \vert_{\tau=\tau_{i+1}(x)}
\cdot \frac{\partial \tau_{i+1}(x)}{\partial x^{\rho}} -
A_{\mu}(c) \frac {\partial c^{\mu}}{\partial \tau} \vert_{\tau=
\tau_{i}(x)} \cdot \frac {\partial \tau_{i}(x)}{\partial x^{\rho}}
+D
\end{equation}
where
\begin{eqnarray}
D=\int_{\tau_{i}}^{\tau_{i+1}} [A_{\mu,\alpha} \cdot \frac{\partial
c^{\alpha}}{\partial x^{\rho}} \cdot \frac{\partial c^{\mu}}
{\partial \tau} + A_{\mu} \frac{\partial^{2} c^{\mu}}
{\partial x^{\rho} \partial \tau} ] d \tau = \nonumber  \\
= A_{\mu}(c) \frac{\partial c^{\mu}}{\partial x^{\rho}} \vert_{\tau
=\tau_{i}}^{\tau=\tau_{i+1}} - \int_{\tau_{i}}^{\tau_{i+1}}
F_{\nu \mu}(c) \frac {\partial c^{\mu}}{\partial x^{\rho}} \cdot
\frac {\partial c^{\nu}}{\partial \tau} d \tau
\end{eqnarray}
Hence
\begin{equation}
\frac{\partial I_{i}}{\partial x_{\rho}} = A_{\mu}(c)
\frac{\partial c^{\mu}}{\partial x^{\rho}}
\vert_{c=c(x,x_{0}),\tau_{i}(x))}^{c=c(x,x_{0},\tau_{i+1}(x))}
- \int_{\tau_{i}}^{\tau_{i+1}} F_{\nu \mu}
\frac{\partial c^{\mu}}{\partial x^{\rho}} \cdot
\frac{\partial c^{\nu}}{\partial \tau} d \tau
\end{equation}
Therefore,
\begin{equation}
\frac{\partial \varphi_{c(x,x_{0})}}{\partial x^{\rho}} =
A_{\mu}(c) \cdot \frac {\partial c^{\mu}}{\partial x^{\rho}}
\vert_{c=c(x,x_{0},0)}^{c=c(x,x_{0},1)} - \int_{0}^{1}
F_{\nu \mu} \frac {\partial c^{\mu}}{\partial x^{\rho}}
\frac {\partial c^{\nu}}{\partial \tau} d\tau
\end{equation}
Using eqn.(\ref{e10}), (\ref{e14}) and (\ref{e15}) we get
\begin{equation}
d \varphi_{c(x,x_{0})}[A] = \left [A_{\rho}(x) - \int_{0}^{1} F_{\nu \mu}(c)
\frac{\partial c^{\mu}}{\partial x^{\rho}}
\frac{\partial c^{\nu}}{\partial \tau} \right] dx^{\rho}
\label{jn1}
\end{equation}
and from $d \varphi = 0$ eqn.(\ref{e18}) follows. Moreover, it can
be directly checked that
\begin{equation}
f_{\mu, \nu}^{c}(x) - f_{\nu, \mu}^{c}(x)=F_{\nu \mu}(x)
\end{equation}
for $f^{c}$ given by eqn.(\ref{e18}) provided that $F_{\mu \nu}$
is antisymmetric and satisfies the Bianchi identities
and that c has to satisfy the
regularity conditions i), ii), iii). \\

It seems rewarding that our gauge condition (\ref{e16}) leads to
the expression (\ref{e18}) for $A_{\mu}$, identical with a known
illustration of the Poincare Lemma \cite{l6,l7}.
In this context it is
worth noticing that the converse of Theorem 1 is true and follows
from eqn.(\ref{jn1}) i.e. if we choose A in eq.(\ref{jn1}) as $f^{c}$
given by eqn.(\ref{e18}), then $d \varphi_{c} [f^{c}] = 0$ i.e.
eqn.(\ref{e16}) is satisfied.\\
Let us remark that we tacitly demand that $V_{4}$ be contractible
\cite{l7} to the point $x_{0}\in V_{4}$ with the admissible
deformations (homotopy) $\zeta$ defined through the conditions  i), ii),
iii) (of course $C^{(\infty)}$ - class homotopy satisfies our
conditions, too).
Therefore, both a set D of points $x_{2}$ to which $V_{4}$ is
$\zeta$ - contractible and sets of admissible curve families
$c^{\mu}(x, x_{2}=const, \tau)$ for each $x_{2} \in D$ are determined
by $V_{4}$ itself. Let us label the relevant collection of all such
$c(x, x_{2} \in D)$
by $\zeta(D)$. \\
To any $c \in \zeta$ there corresponds a unique electromagnetic
potential $A=f^{c}$ satisfying eqn.(\ref{e16}). Let us check when
different curves may correspond to the same $f$. Take two different
curve families $c_{1}(x, x_{01})$, $c_{2}(x, x_{02})$ and assume
\begin{equation}
f^{c_{1}}=f^{c_{2}}
\label{jn3}
\end{equation}
\begin{equation}
i.e.\;\; d \varphi_{c_{1}}[f^{c_{1}}]=d \varphi_{c_{2}}[f^{c_{1}}]=0
\label{jn4}
\end{equation}
First, notice that
\begin{equation}
d \varphi_{c}[A]=0 \Longleftrightarrow \varphi_{c(x, x_{0})}=
\varphi_{c(x_{0},
x_{0})}=c[F]
\label{jn5}
\end{equation}
where
\begin{equation}
c[F]=\oint_{c(x_{0},x_{0})} A dc
\label{jn6}
\end{equation}
where $c[F]$ does not depend
on $x$ and $c[F] \neq 0$ unless $c(x,x_{0})$
is contractible to a tree \cite{l8} i.e. to a closed path (e.g.
a point) with "null area". \\
Next, from (\ref{jn4}) and (\ref{jn5}) it follows that
\begin{equation}
\int_{c_{1}(x, x_{01})\bar{c}_{2}(x,x_{02})}f^{c_{1}}(y) dy =
c_{1}[F] - c_{2}[F]
\label{jn7}
\end{equation}
for any $x$ and
therefore,
\begin{equation}
\oint_{L} f^{c_{1}}(y) dy = 0
\label{jn8}
\end{equation}
for any $x$, $x' \in V_{4}$ ;
\begin{equation}
L=c_{1}(x,x_{01})\bar{c}_{2}(x,x_{02}) c_{2}(x',x_{02})\bar{c}_{1}
(x',x_{01})
\label{jn9}
\end{equation}
As eqn.(\ref{jn8}) has to be true for any chosen electromagnetic,
$C^{(1)}$- regular field $F$, $L$ has to be a tree in order to satisfy
(\ref{jn8}). Therefore we should consider a set of classes $\zeta(D)/T$
with equivalence relation $T$:
\begin{quotation}
\begin{equation}
c_{1}(x,x_{01})=c_{2}(x,x_{02})
\label{jn10}
\end{equation}
if, for any $x$, $x' \in V_{4}$ the closed path $L(c_{1}\bar{c}_2
c_{2}^{,}\bar{c}_{1}^{,})$ (eqn. \ref{jn9}) is a tree \cite{l8}.
\end{quotation}
(It can be checked that relation (\ref{jn10}) is
an equivalence relation)
Choosing a single representative for each class $c/T$
we get a one to one
correspondence $c \longleftrightarrow f^{c}$. The permutation of
a set $ \zeta(D)/T$
\begin{equation}
P: c \longrightarrow P(c)
\label{jn11}
\end{equation}
implies
\begin{equation}
f^{c} \longrightarrow P f^{c}=f_{P(c)}
\label{jn12}
\end{equation}
which, after the use of eqn.(\ref{e18}) and (\ref{jn1}) (inserting
$A=f^{c}$)reads
\begin{equation}
f^{c}(x) dx \longrightarrow f^{P(c)}(x) dx = f^{c}(x) dx - d \varphi_
{P(c)}[f^{c}]
\label{jn14}
\end{equation}
From eqns.(\ref{e18}),(\ref{jn1}), we easily find that
\begin{equation}
d \varphi_{c_{1}}[f^{c_{2}}] + d \varphi_{c_{2}}[f^{c_{1}}] = 0
\label{jn15}
\end{equation}
so that (\ref{jn14}) becomes
\begin{equation}
f^{c}(x) dx = f^{P(c)}(x) dx - d \varphi_{c}[f^{P(c)}]
\label{jn16}
\end{equation}
Finally, comparing (\ref{jn14}) with a familiar form
\begin{equation}
f^{P(c)} dx = f^{c} dx - \frac{i}{e} ~\omega_{c \rightarrow P(c)}^{-1} d \omega
_{c \rightarrow P(c)}
\label{jn17}
\end{equation}
we get
\begin{equation}
d \varphi_{P(c)}[f_{c}]= \frac{i}{e} d \ln \omega_{c \rightarrow P(c)}
\label{jn18}
\end{equation}
i.e.
\begin{equation}
\omega_{c \rightarrow P(c)} = \omega_{0} \exp( - i e
\varphi_{P(c)}[f_{c}])
\label{jn19}
\end{equation}
Let us add, that the transition $A \longrightarrow f^{c}$ can be
obtained similarly for an arbitrary potential A:
\begin{equation}
f^{c}=A - \frac{i}{e} \omega^{-1}(A \rightarrow f^{c}) \partial
\omega(A \longrightarrow f^{c})
\label{jn20}
\end{equation}
with
\begin{equation}
\omega(A \longrightarrow f^{c}) = \omega_{0} \exp(-i e \varphi_{c}[A])
\label{jn21}
\end{equation}
so that any $A$ can be gauge transformed to $f^{c}$.

\section{Yang-Mills theory}
\setcounter{equation}{0}
The generalization of the above results to Yang-Mills fields is
relatively straightforward except for one complication linked
to the Bianchi identity.
In the Yang-Mills theories the field $G_{\mu \nu}$ is defined as
\begin{equation}
G_{\mu \nu}=F_{\mu \nu} + i g [A_{\mu}, A_{\nu}]
\label{at1}
\end{equation}
where
\begin{equation}
F_{\mu \nu}=\partial_{\mu}A_{\nu} -
\partial_{\nu} A_{\mu}
\label{at2}
\end{equation}
and
\begin{equation}
A_{\mu}=A_{\mu}^{a} T_{a}  \nonumber
\end{equation}
From (\ref{at1}), (\ref{at2}) the Bianchi identities follow:
\begin{equation}
D_{\mu}G_{\nu \rho} + D_{\nu}G_{\rho \mu} + D_{\rho}G_{\mu \nu} = 0
\label{at3}
\end{equation}
where
\begin{equation}
D_{\mu} = \partial_{\mu} + i g [A_{\mu}, \cdot ]
\label{at4}
\end{equation}
The condition (\ref{e16}) applied to the matrix $A_{\mu}$:
\begin{equation}
d \varphi_{c(x, x_{0})}[A] = 0
\label{at5}
\end{equation}
for a curve with
\begin{equation}
c^{\mu}(x^{0},x^{0})= x^{0^{\mu}}
\label{at6}
\end{equation}
leads - due to eqns. (\ref{jn5}),(\ref{jn6}) - to
\begin{equation}
\varphi_{c(x,x_{0})}=0
\label{at7}
\end{equation}
Therefore, trivially
\begin{equation}
\partial_{\mu} \varphi = D_{\mu} \varphi = 0
\label{at8}
\end{equation}
Theorem 1 can be applied to $A$ with a given $F$ and, using (\ref{e18})
with (\ref{at1}) we get
\begin{equation}
f_{\rho}^{c}= \int_{c} \left(G_{\nu \mu}(c) - i g [f_{\nu}^{c},
f_{\mu}^{c}] \right) \frac{ \partial c^{\mu}}{\partial x^{\rho}}
\frac{ \partial c^{\nu}}{\partial \tau} d \tau
\label{at9}
\end{equation}
In order to simplify (\ref{at9}) let us limit ourselves to a class
of curve families satisfying as explained below a self-contractibility
condition:
\begin{center}
For any $y \in c(x, x_{0})$,
\end{center}
\begin{equation}
c(x, x_{0}) \cap c(y, x_{0}) = c(y
, x_{0})
\label{at10}
\end{equation}
This condition means that the curves defined by:
\begin{equation}
y \in \tilde{c}(c(x, x_{0},t), x_{0}) \Longleftrightarrow y^{\mu} =
c^{\mu}(x, x_{0}, \tau_{1})
\label{at11}
\end{equation}
for some $0 \leq \tau_{1} \leq t$ \\
and
\begin{equation}
y \in c(c(x, x_{0}, t), x_{0}) \Longleftrightarrow y^{\mu} =
c^{\mu}(c(x, x_{0}, t), x_{0}, \tau)
\label{at12}
\end{equation}
for some $0 \leq \tau \leq 1$ \\
are the same for any $0 \leq t \leq 1$. Therefore, there must exist
a change of parametrization
\begin{eqnarray}
\tau \longrightarrow h(\tau, x, t), \;\ \frac{\partial h}{\partial \tau}
> 0 \\
h(1, x, t) = t \nonumber \\
h(0, x, t) = 0 \nonumber
\label{at14}
\end{eqnarray}
leading to the following relation
\begin{equation}
c^{\mu}(c(x, x_{0}, t), x_{0}, \tau) = c^{\mu}(x, x_{0}, h(\tau, x, t))
\label{at15}
\end{equation}
We are now in a position to prove the following theorem:

\begin{th1}
For any self-contractible family of curves, c, eqn.(\ref{at1})
together with the condition (\ref{at5}) are equivalent to:
\begin{equation}
A_{\rho} = f_{\rho}^{c}(x) \equiv \int_{c(x, x_{0})} G_{\nu \mu}(y)
\frac{\partial y^{\mu}}{\partial x^{\rho}} d y^{\nu}
\label{t15a}
\end{equation}
provided that $f^{c}[G]$, $G$ satisfy the Bianchi identities (\ref{at3}).
\end{th1}
{\it \underline{Proof}}\\
Firstly let us prove that (\ref{t15a}) follows from (\ref{at1}),
(\ref{at5}).
This will be achieved by the proof of
the following lemma.
\\
{\bf Lemma}\\
For any self-contractible curve family c, and any antisymmetric
$T_{\mu \nu}$, a vector field $f_{\rho}^{c}(x)$ defined as
\begin{equation}
f_{\rho}^{c}(x) = \int_{c(x,x_{0})} T_{\nu \mu}(y)
\frac {\partial y^{\mu}}{\partial x^{\rho}} d y^{\nu}
\label{t15b}
\end{equation}
is perpendicular to the curves
of this family:
\begin{eqnarray}
\frac{ \partial c^{\rho}(x, x_{0}, t)}{\partial t} f_{\rho}^{c}
(c^{\mu}(x, x_{0}, t)) = 0
\label{t151} \\
0 \leq t \leq 1   \nonumber
\end{eqnarray}
{\it \underline{Proof of lemma}}\\
Let
\begin{equation}
R=\frac{ \partial c^{\rho}(x, x_{0}, t)}{\partial t} f_{\rho}^{c}
(c^{\mu}(x, x_{0}, t))
\label{at16}
\end{equation}
Using (\ref{t15b}) we get
\begin{eqnarray}
R = \frac{\partial c^{\rho}(x, x_{0}, t)}{\partial t} \int T_{\nu \mu}
(c(c(x, x_{0}, t), x_{0}, \tau)) \nonumber \\
\frac{\partial c^{\mu}(c(x, x_{0}, t), x_{0}, \tau)}{\partial c^{\rho}
(x, x_{0}, t)} \frac{ \partial c^{\nu}(c, x_{0}, \tau)}{\partial \tau}
d \tau
\label{at17}
\end{eqnarray}
Hence
\begin{equation}
R = \int T_{\nu \mu}(c(c, x_{0}, \tau)) \frac{\partial c^{\mu}
(c(x, x_{0}, t), x_{0}, \tau)}{\partial t}
\frac{\partial c^{\nu}(c, x_{0}, \tau)}{\partial \tau} d \tau
\label{at18}
\end{equation}
and using (\ref{at15}) we get
\begin{eqnarray}
R = \int T_{\nu \mu}(c(x, x_{0}, h(\tau, x, t))) \frac{\partial c^{\mu}
(x, x_{0}, h)}{\partial t} \frac{\partial c^{\nu}(x, x_{0}, h)}
{\partial \tau} d \tau =  \nonumber \\
= \int T_{\nu \mu} \frac{\partial c^{\mu}}
{\partial h} \frac{\partial c^{\nu}}{\partial h} \frac{\partial h}
{\partial t} \frac{\partial h}{\partial \tau} d \tau
\label{at19}
\end{eqnarray}
i.e.
\begin{equation}
R = 0
\label{at20}
\end{equation}
due to the antisymmetry of $T_{\mu \nu}$ in (\ref{at19}). \\
Inserting eqn.(\ref{t151}) into (\ref{at9}) we get
\begin{equation}
f_{\rho}^{c} = \int G_{\nu \mu}(c) \frac{\partial c^{\mu}}{\partial
x^{\rho}} \frac{\partial c^{\nu}}{\partial \tau} d \tau
\label{at21}
\end{equation}
for any self-contractible family of curves $c$ i.e. we get eqn.(
\ref{t15a}) of theorem 2.
Eqn.(\ref{at3}) trivially follows from (\ref{at1}) for any $A$.
The second part of the proof of
Theorem 2 is given in Appendix A; we show there, that $f^{c}[G]$
satisfies eqns.(\ref{at1}), (\ref{at5}) if (\ref{at3})
is fulfilled. \\

For $c \longrightarrow c^{,}$ ($c$, $c^{,}$ self-contractible)
\begin{equation}
f^{c^{,}}= \omega_{c \rightarrow c^{,}}^{-1} f^{c}
\omega_{c \rightarrow c^{,}} - \frac{i}{g} \omega_{c \rightarrow c^{,}}
^{-1} \partial \omega_{c \rightarrow c^{,}}
\label{at22}
\end{equation}
with
\begin{equation}
\omega_{c \rightarrow c^{,}} =  P \exp \left(-ig \int_{0}^{1}
f^{c}(c^{,}) \frac{\partial c^{,}}{\partial \tau} d \tau \right) \omega_{0} 
\label{at23}
\end{equation}
\underline{Proof}: Take any $x = c^{,}(x, x_{0}, t)$.then, from
(\ref{t151}):
\begin{equation}
0=\omega_{c \rightarrow c^{,}}^{-1} f_{\rho}^{c}(c^{,}(x, x_{0}, t))
\omega_{c \rightarrow c^{,}} \frac{\partial c^{ \rho^{,}}}{\partial t}
- \frac{i}{g} \omega_{c \rightarrow c^{,}}^{-1} \frac{\partial c^{,}}
{\partial t} \frac{\partial}{\partial c^{,}} \omega_{c}
\label{at24}
\end{equation}
i.e.
\begin{equation}
f_{\rho}^{c}(c^{,}(x, x_{0}, t))
\frac{\partial c^{,^{\rho}}}{\partial t}
\omega_{c \rightarrow c^{,}} = \frac{i}{g} \frac{\partial}{\partial t}
\omega_{c \rightarrow c^{,}}
\label{at25}
\end{equation}
This is solvable, and for $c^{,}(x, x_{0}, 1) \equiv x$ gives
(\ref{at23}). Moreover (\ref{at23}) remains true for any
$A \longrightarrow f^{c^{,}}$ transition, showing that any $A$
can be gauge transformed into $f^{c}$.

\section{Relation to other Yang-Mills gauges}
\setcounter{equation}{0}
We show now that
those gauge conditions utilized \cite{l3,l4,l5} in Yang-Mills theory
which lead to the form $A=A[G]$; are special case of our ponderomotive
gauges coresponding to special cases of the curves c.
If we take $c(0, x^{\mu}) = s(0, x^{\mu})$, evidently
satisfying condition (\ref{at10}) we get from (\ref{t151})
\begin{equation}
A_{\mu}x^{\mu} = 0
\label{at26}
\end{equation}
i.e. the Fock-Schwinger gauge \cite{l3}. \\
For $c(0, x^{\mu}) = s(0, [0, \vec{x}]) s([0,\vec{x}],[x^{0},\vec{x}]$,
where $[a, \vec{b}]$ denote a space-time point,
we get temporal and space-like Fock-Schwinger \cite{l5} gauges:
\begin{eqnarray}
\vec{A}(0, \vec{x}) \cdot \vec{x} = 0   \nonumber \\
A^{0}(x_{0}, \vec{x}) = 0
\label{at27}
\end{eqnarray}
Finally, choosing
\begin{eqnarray}
c(0, x^{\mu}) = s(0, [x^{0}, \vec{0}]) s([x^{0}, \vec{0}],
[x^{0}, x^{1},0,0]) s([x^{0}, x^{1}, 0, 0],
[x^{0}, x^{1}, x^{2}, 0])  \nonumber \\
s([x^{0}, x^{1}, x^{2}, 0],
[x^{0}, x^{1}, x^{2}, x^{3}])
\label{at28}
\end{eqnarray}
we get the superaxial gauge (see e.g.\cite{l4}):
\begin{eqnarray}
A_{0}(x^{0}, \vec{0}) = 0 \;\;\;\;\ A_{2}(x^{0}, x^{1}, x^{2}, 0) = 0
\nonumber \\
A_{1}(x^{0}, x^{1}, 0, 0) = 0 \;\;\;\;\ A_{3}(x^{\mu})=0
\label{at29}
\end{eqnarray}

\section{Path dependent Green's functions}
\setcounter{equation}{0}
Now we consider a Dirac particle interacting with a given
electromagnetic field. Use of our ponderomotive gauges enables us
to generalize Schwinger's method of extracting a gauge invariant
factor from the fermion propagator and leads to a perturbative
expansion for the propagator in terms of the electromagnetic
field itself. We show that Schwingers approach to this problem could
lead to incorrect results. \\
Starting with the Dirac equation
\begin{equation}
(\der + i e \Akr +im) \Psi(A\mid x) = 0
\label{cz1}
\end{equation}
we substitute, for some family $c(x, x_{0})$:
\begin{equation}
\Psi(f^{c} \mid x) = \omega(A \longrightarrow f^{c}) \Psi(A \mid x)
\label{cz2}
\end{equation}
where (compare eqn.(\ref{jn21}); we put $\omega_{0} = 1$):
\begin{equation}
\omega(A \longrightarrow f^{c}) = \exp (-ie \varphi_{c}[A])
\label{cz3}
\end{equation}
Then, using (\ref{jn20}) we get
\begin{equation}
(\der + i e \fkr ^{c} +i m) \Psi(\fkr ^{c} \mid x) = 0
\label{cz4}
\end{equation}
Similarly, the Green function
\begin{equation}
G(A \mid x_{1}, x_{2}) = -i <0 \mid T(\Psi(A \mid x_1)
\bar{\Psi}(A \mid x_{2})) \mid 0 >
\label{cz5}
\end{equation}
satisfying
\begin{equation}
(\der_{1} + i e \Akr + i m) G(A \mid x_{1}, x_{2}) = -i \delta(x_{1}
- x_{2})
\label{cz6}
\end{equation}
\begin{equation}
G(A \mid x_{1}, x_{2}) (-\stackrel{\leftarrow}{\der}_{2} + i e
\Akr(x_{2}) + i m) = - i \delta(x_{2} - x_{1})
\label{cz7}
\end{equation}
can be replaced by
\begin{equation}
G(f^{c} \mid x_{1}, x_{2}) = \omega(A \longrightarrow f^{c(x_{1},
x_{0})}) G^{A} \omega^{-1}(A \longrightarrow f^{c(x_{2}, x_{0})})
\label{cz8}
\end{equation}
and eqns.(\ref{cz6}), (\ref{cz7}) are replaced by
\begin{equation}
(\der_{1} + i e \fkr^{c}(x_{1}) + i m) G(f^{c} \mid x_{1}, x_{2})
= - i \delta(x_{1} - x_{2})
\label{cz9}
\end{equation}
\begin{equation}
G(f^{c} \mid x_{1}, x_{2}) (- \stackrel{\leftarrow}{\der}_{2} + i e
\fkr^{c}(x_{2}) + i m) = - i \delta (x_{1} - x_{2})
\label{cz10}
\end{equation}
They have the same iterative solution as (\ref{cz6}), (\ref{cz7}):
\begin{equation}
G(f^{c} \mid x_{1}, x_{2}) = S + e S\fkr_{c}S + e^{2} S\fkr_{c}S
\fkr_{c}S + \ldots
\label{cz11}
\end{equation}
where
\begin{equation}
S(x, y) = \int \frac{d^{4}p}{(2 \pi)^{4}} \frac{\pkr + m}{p^{2} -
m^{2} + i\varepsilon} e^{-ip(x-y)}
\label{cz12}
\end{equation}
The above suggests that an approach used by Schwinger \cite{l2}
may not be generally applicable.
Schwinger uses a different transformation from ours (\ref{cz8})
namely:
\begin{equation}
\widetilde{G}(x_{1}, x_{2}) = \omega(A \longrightarrow f^{c(x_{1},
x_{2})}) G(A \mid x_{1}, x_{2})
\label{cz13}
\end{equation}
Although the definitions (\ref{cz8}) and (\ref{cz13}) differ only
by a (known) phase factor, the difference could become important
in calculations involving derivatives of the Green's function.
While in eqns.(\ref{cz9}), (\ref{cz10}) involving \\
$G(f_{c} \mid x_{1},
x_{2})$ only one function
\begin{equation}
f_{(x, x_{0})}^{c,L} \equiv f^{c}(x, x_{0})
\label{cz14}
\end{equation}
occurs in both equations, the equations for $\widetilde{G}$
will involve two functions, $f^{c, L}$ and $f^{c, R}$:
\begin{equation}
(\der_{1} + i e \fkr^{c, L}(x_{1}, x_{2}) + i m) \widetilde{G}(x_{1},
x_{2}) = - i \delta(x_{1} - x_{2})
\label{cz15}
\end{equation}
\begin{equation}
\widetilde{G}(x_{1}, x_{2})(- \stackrel{\leftarrow}{\der}_{2} +
i e \fkr^{c, R}(x_{1}, x_{2}) + i m) = - i \delta(x_{1}-x_{2})
\label{cz16}
\end{equation}
where
\begin{equation}
f_{\rho}^{c, L}(x_{1}, x_{2}) = \int_{c(x_{1}, x_{2})} F_{\nu \mu}(x)
\frac{\partial x^{\mu}}{\partial x_{1}^{\rho}} d x^{\nu}
\label{cz17}
\end{equation}
\begin{equation}
f_{\rho}^{c, R}(x_{1}, x_{2}) = \int_{c(x_{1}, x_{2})} F_{\nu \mu}(x)
\frac{\partial x^{\nu}}{\partial x_{2}^{\rho}} d x^{\mu}
\label{cz18}
\end{equation}
Both these functions are potentials corresponding to the same
$F_{\mu \nu}$ i.e.
\begin{equation}
\frac{\partial}{\partial x_{1}^{\sigma}} f_{\rho}^{c, L}(x_{1}, x_{2})
- \frac{\partial}{\partial x_{2}^{\rho}} f_{\sigma}^{c, L}(x_{1},
x_{2}) = F_{\sigma \rho}(x_{1})
\label{cz19}
\end{equation}
\begin{equation}
\frac{\partial}{\partial x_{2}^{\sigma}} f_{\rho}^{c, R} (x_{1}, x_{2})
- \frac{ \partial}{\partial x_{2}^{\rho}} f_{\sigma}^{c, R}(x_{1},
x_{2}) = F_{\sigma \rho}(x_{2})
\label{cz20}
\end{equation}
but, in general, they are not identical:
\begin{equation}
f_{\mu}^{c, R}(x_{1}, x_{2}) \neq f_{\mu}^{c, L}(x_{1}, x_{2})
\label{cz21}
\end{equation}
This casts some doubt on the general applicability of Schwinger
replacement \cite{l2}
\begin{equation}
\widetilde{G}(x_{1}, x_{2}) \longrightarrow <x_{1} \mid \widetilde{G}
\mid x_{2}>
\label{cz22}
\end{equation}
with $\widetilde{G}$ treated as the same operator in both eqns.
(\ref{cz15}) and (\ref{cz16}).

\section{Ponderomotive gauges in the canonical formalism - Maxwell
theory}
\setcounter{equation}{0}
We consider now the use of our ponderomotive gauges in the quantized
field theoretic context i.e. in QED. We shall show that our gauge
constraint and Gauss's law are canonically conjugate leading to
a rather simple form for the Dirac brackets of the theory.
The canonical Hamiltonian in Maxwell theory
\begin{equation}
H_{c} = \int d^{3}x \left[ \frac{1}{2} \vec{\Pi}^{2} + \frac{1}{2}
\vec{B}^{2} + A^{0} \, \vec{\nabla} \cdot \vec{\Pi} \right]
\label{pi1}
\end{equation}
with the primary constraints:
\begin{equation}
D_{1} = \Pi^{0} = 0
\label{pi2}
\end{equation}
\begin{equation}
- D_{2} = \partial_{i} \Pi_{i} = 0
\label{pi3}
\end{equation}
requires two futher constraints before the construction
of the Dirac brackets
\cite{l9} \\
Let us choose them to be a temporal gauge condition
\begin{equation}
D_{3} = A_{0} = 0
\label{pi4}
\end{equation}
and ponderomotive space-like gauge condition
\begin{equation}
D_{4} = \int_{0}^{1} A_{i}(\vec{c}(\vec{x}, \tau)) \frac{\partial
c_{i}}{\partial \tau} d \tau = 0
\label{pi5}
\end{equation}
at a given moment of time. The curve c runs from $\vec{c}(\vec{x}, 0)
= \vec{x}_{0}$ to $\vec{c}(\vec{x}, 1) = \vec{x}$ \\
The corresponding Hamiltonian
\begin{equation}
H = H_{c} + \int d^{3}x v_{2} \partial_{i} \Pi_{i}
\label{pi6}
\end{equation}
where
\begin{equation}
v_{2} = \int_{0}^{1} \Pi_{i}(c) \frac{\partial c_{i}(\vec{x}, \tau)}
{\partial \tau} d \tau
\label{pi7}
\end{equation}
weakly commutes with the constraints $D_{i}$. \\
The Poisson brackets between the $D_{j}$ turn out to be:
\begin{equation}
c_{31} = - c_{13} \equiv [D_{3}(\vec{x}), D_{1}(\vec{y})]_{P}
= \delta_{3}
(\vec{x} - \vec{y})
\label{pi8}
\end{equation}
\begin{equation}
c_{42}(x,y) = - c_{24}(y,x) \equiv [D_{4}(\vec{x}), D_{2}(\vec{y})]_{P}
= \delta_{3}
(\vec{y} - \vec{x}) - \delta_{3}(\vec{y} - \vec{x}_{0})
\label{pi9}
\end{equation}
The rest of Poisson brackets vanish. As an example
let us derive eqn.(\ref{pi9}). We have
\begin{equation}
c_{42} = [D_{4}(x), D_{2}(y)]_{P} = - \frac{\partial}{\partial y_{k}}
\int_{0}^{1} d \tau \frac{\partial c_{k}}{\partial \tau} \delta(y_{1}
- c_{1}) \delta(y_{2} - c_{2}) \delta(y_{3} - c_{3})
\label{pi10}
\end{equation}
Hence
\begin{eqnarray}
c_{42} = \int_{0}^{1} d \tau \frac{\partial c_{i}}{\partial \tau}
\frac{\partial}{\partial c_{i}} \delta_{3}(\vec{y} - \vec{c}(\vec{x},
\tau)) d \tau   \nonumber \\
= \delta_{3}(\vec{y} - \vec{c}(\vec{x}, 1)) - \delta_{3}(\vec{y} -
\vec{c}(\vec{x}, 0))   \nonumber \\
= \delta_{3}(\vec{y} - \vec{x}) - \delta_{3}(\vec{y} - \vec{x}_{0})
\label{pi11}
\end{eqnarray}
If now we discard the single point $\vec{x} = \vec{y} = \vec{x}_{0}$
the $c_{ik}$ can be written as
\begin{displaymath}
c_{ik}=\left[ \begin{array}{cc}
0 & -I \\
I &  0
\end{array}
\right] \cdot \delta_{3}(\vec{x} - \vec{y})
\label{pi12}
\end{displaymath}
so that
\begin{equation}
c_{ik}^{-1} = - c_{ik}
\label{pi13}
\end{equation}
The simplicity of the $c_{ik}$ stems from the choice of the
ponderomotive
gauge condition $D_{4}$; moreover, and this is very important,
$D_{4}$ and $D_{2}$
(Gauss's law) are canonically conjugate. \\
The Dirac brackets of the theory,
\begin{equation}
[A, B]_{D} = [A, B]_{P} - [A, D_{i}]_{P} \; c_{i k}^{-1} \; [D_{k}, B]_{P}
\label{pi14}
\end{equation}
become in our case
\begin{eqnarray}
[A, B]_{D} = [A, B]_{P} + [A, D_{3}]_{P} \cdot [D_{1}, B]_{P} - \nonumber \\
- [A, D_{1}]_{P} \cdot [D_{3}, B]_{P} + \nonumber \\
+ [A, D_{4}]_{P} \cdot [D_{2}, B]_{P} - [A, D_{2}]_{P} \cdot [D_{4}, B]_{P}
\label{pi15}
\end{eqnarray}
For $\vec{\Pi}$, $\vec{A}$ the Poisson brackets with $D_{i}$'s are:
\begin{equation}
[A_{i}(\vec{x}), D_{2}(\vec{z})]_{P} = \frac{\partial}{\partial x^{i}}
\delta_{3}(\vec{x} -
\vec{z})
\label{pi16}
\end{equation}
\begin{equation}
[\Pi_{i}(\vec{x}), D_{4}(\vec{z})]_{P} = - \int_{0}^{1} \delta_{3}
(\vec{x} - \vec{c}(\vec{z}, \tau)) \frac{\partial c^{i}}{\partial \tau}
d \tau
\label{pi17}
\end{equation}
and the other brackets vanish. \\
So, finally,
for
\begin{eqnarray}
\vec{x} \neq \vec{x}_{0} \nonumber \\
\vec{y} \neq \vec{y}_{0}
\label{pi18}
\end{eqnarray}
\begin{eqnarray}
[\Pi_{i}(x), A_{j}(y)]_{D} = - \delta_{ij} \delta(\vec{x} - \vec{y}) +
[\Pi_{i}, D_{4}]_{P} \cdot [D_{2}, A_{j}]_{P}  \nonumber \\
= - \delta_{ij} \delta(\vec{x} - \vec{y}) + \int d^{3} z \int d \tau
\delta_{3}(\vec{x} - \vec{c}(\vec{z}, \tau)) \frac{\partial c^{i}}
{\partial \tau} \partial_{j} \delta_{3}(\vec{y} - \vec{z}) \nonumber \\
= - \delta_{ij} \delta(\vec{x} - \vec{y}) + \frac{\partial}
{\partial y_{j}} \int d \tau \delta_{3}(\vec{x}- \vec{c}(\vec{y},\tau))
\frac{\partial c^{i}}{\partial \tau} d \tau
\label{pi19}
\end{eqnarray}
The other Dirac brackets vanish. $\Pi^{0}$ and $A^{0}$ being
constraints, have vanishing Diracs brackets with any variable.

\section{Dirac brackets for Y-M theory}
\setcounter{equation}{0}

In the case of Yang-Mills theory we limit ourselves to self-contractible
families, defined by eqn.(\ref{at10}). They have a useful property established earlier
(compare Lemma in the proof of Theorem 2),
namely Y-M potentials are orthogonal to these curves, i.e. from
the gauge constraints
\begin{equation}
\int_{c(x,x_{0})} A_{a}^{\mu}(y) dy_{\mu} = 0
\label{j1}
\end{equation}
follows eqn.(\ref{t151})
\begin{equation}
\frac{\partial c_{\mu}(x, x_{0}, \tau)}{\partial \tau} A_{a}^{\mu}
(c(x, x_{0}, \tau)) = 0
\label{j2}
\end{equation}
for any $0 \leq \tau \leq 1$ and $x \in V$. This relation will allow us to establish eq.(\ref{d12})
which in turn is a crucial condition responsible for the simple form of Dirac brackets.\\
We are going to implement these gauges into canonical
formalism of Y-M theory. \\
In what follows the discussion of surface terms will be omitted.
The canonical Hamiltonian is then:
\begin{equation}
H = \int_{V} d^{3}x {\cal H}
\label{d1}
\end{equation}
with
\begin{eqnarray}
{\cal H} = \frac{1}{2} (\vec{B}_{a} \cdot \vec{B}_{a} + \vec{E}_{a}
\cdot \vec{E}_{a}) -
[\vec{\nabla} \cdot \vec{E}_{a} - g {\cal C}_{abc} \vec{A}_{b} \cdot
\vec{E}_{c}] A_{a}^{0}
\label{d2}
\end{eqnarray}
\vspace{0.25cm}
\begin{equation}
E_{a}^{i}(\vec{x}) = - \Pi_{a}^{i}(\vec{x})
\label{d3}
\end{equation}
\begin{equation}
D_{a}^{(1)} = \Pi_{a}^{0} \approx 0
\label{d4}
\end{equation}
\begin{equation}
D_{a}^{(2)} = \vec{\nabla} \cdot \vec{E}_{a} - g
{\cal C}_{abc} \vec{A}_{b} \cdot \vec{E}_{c}
\approx 0
\label{d5}
\end{equation}
We take temporal gauge
\begin{equation}
D_{a}^{(3)} = A_{a}^{0} \approx 0
\label{d6}
\end{equation}
and ponderomotive space-like gauge constraint
\begin{equation}
D_{a}^{(4)} = \int_{c(\vec{x}, \vec{x}_{0})} A_{a}^{i}(y)
dy^{i} \approx 0
\label{d7}
\end{equation}
The constraints (\ref{d4}-\ref{d7}) are compatible with
\begin{equation}
{\cal H}' = \frac{1}{2}(\vec{B}_{a} \cdot \vec{B}_{a} + \vec{E}_{a}
\cdot \vec{E}_{a}) + D_{a}^{(2)}(x) v_{(a)}^{(2)}(x)
\label{d8}
\end{equation}
where
\begin{equation}
v_{a}^{(2)}(x) = \int_{c(\vec{x}, \vec{x}_{0})} dy^{i}
E_{a}^{i}(y)
\label{d9}
\end{equation}
Our further
considerations will be valid for the region $V_{-}$:
\begin{equation}
V_{-} = V - P(\vec{x}_{0})
\label{d10}
\end{equation}
Next, let us remark that compatibility of (\ref{d7}) with (\ref{d8})
is evident once we prove - in analogy with Maxwell theory - that
in $V_{-}$
\begin{equation}
[D_{d}^{(4)}(\vec{x}), D_{a}^{(2)}(\vec{y})]_{P} =
\delta(\vec{x} - \vec{y})
\delta_{ad}
\label{d11}
\end{equation}
The first term of $D_{a}^{(2)}(y)$, $- \vec{\nabla}\cdot \vec{E}_{a}$
(comp. eq.(\ref{d5}))yields already r.h.s. of (\ref{d11}) - derivation
is the same as for Maxwell theory. So we have to
show that
\begin{equation}
{\cal C}_{abc} [D_{a}^{(4)}(\vec{x}), A_{b}^{i} E_{c}^{i}(y)]_{P}
\approx 0
\label{d12}
\end{equation}
The use of (\ref{d7}) gives
\begin{eqnarray}
{\cal C}_{abc} [D_{a}^{(4)}(x), A_{b}^{i}(y) E_{c}^{i}(y)]_{P}= \nonumber \\
= {\cal C}_{abc} \int_{c(\vec{x}, \vec{x}_{0})} dz^{k} [A_{d}^{k}(z),
A_{b}^{i}(y)E_{c}^{i}(y)]_{P}= \nonumber \\
= {\cal C}_{abc} (-) \delta_{cd} \int_{c(\vec{x}, \vec{x}_{0})} dz^{k}
\delta(\vec{z}-\vec{y}) A_{b}^{k}(z)
\label{d13}
\end{eqnarray}
Please notice, that $dz^{k} A^{k}(z)|_{z\in c(\vec{x}, \vec{x}_{0})}
\approx 0$ from (\ref{d7}) (comp.eqns (\ref{j1}), (\ref{j2})),
therefore (\ref{d12}) is proved.\\
Let us come back to constraints ((\ref{d4})-(\ref{d7})). With the
help of (\ref{d11})the matrix
\begin{equation}
d_{a, b}^{i, k} = [D_{a}^{(i)}(x), D_{b}^{(k)}(y)]_{P}
\label{d14}
\end{equation}
can be written as
\begin{equation}
d_{a, b}^{i, k}=\left[ \begin{array}{cc}
0 & -I \\
I &  0
\end{array}
\right]^{ik} \cdot \delta_{ab} \delta_{3}(\vec{x} - \vec{y})
\label{d15}
\end{equation}
so that
\begin{equation}
d^{-1} = - d
\label{d16}
\end{equation}
and Dirac brackets of the theory follow \cite{l9}:
\begin{eqnarray}
[E_{c}^{r}(\vec{x}), A_{d}^{s}(\vec{y})]_{D} = \delta_{cd} \delta_{rs}
\delta(\vec{x}-\vec{y}) - \nonumber \\
- \left[ \frac{\partial}{\partial y^{s}} \delta_{cd}
- g {\cal C}_{cdb} A_{b}^{s}(y) \right]
\cdot \int_{w \in c(\vec{y}, \vec{x}_{0})} dw^{r} \delta(\vec{x}-\vec{w})
\label{d17}
\end{eqnarray}
\vspace{0.35cm}
\begin{equation}
[A_{c}^{r}, A_{d}^{s}]_{D} = 0
\label{d18}
\end{equation}
\vspace{0.27cm}
\begin{eqnarray}
[E_{c}^{r}(\vec{x}), E_{d}^{s}(\vec{y})]_{D} = \nonumber \\
= g {\cal C}_{cdf}
\left[ \int_{w\in c(\vec{x}, \vec{x}_{0})} dw^{s} \delta(\vec{y} -
\vec{w}) E_{f}^{r}(\vec{x})
+ \int_{w \in c(\vec{y}, \vec{x}_{0})} dw^{r} \delta(\vec{x} -
\vec{w}) E_{f}^{s}(y) \right]
\label{d19}
\end{eqnarray}
In the next section eqns.(\ref{d17} - \ref{d19}) will be used in the derivation
of nonabelian algebras.

\section{Flux algebras}
\setcounter{equation}{0}

In this chapter we are going to establish algebras
of fluxes:
\begin{equation}
{\cal B}_{a}^{(\sigma)} = \int_{\sigma} B_{a}^{k} n^{k}
d\sigma
\label{j21}
\end{equation}
\begin{eqnarray}
{\cal E}_{a}^{(\sigma^{*})} = \int_{\sigma^{*}}
E_{a}^{k} n^{k} d
\sigma
\label{j3} \\
\vec{x}^{0} \inkr \sigma  \nonumber
\end{eqnarray}
The surfaces $\sigma$, $\sigma^{*}$ are not interrelated. In the first part of this chapter
we shall choose these surfaces in such a manner that the fluxes ${\cal B}$, ${\cal E}$ will be 
equal to loop integrals over the potential and dual potential, respectively. \\
Let us consider at the beginning a special type of surfaces
appearing in definitions (\ref{j21}), (\ref{j3})
of ${\cal B}$, ${\cal E}$
fluxes. Take a loop $L$ and some homotopy $c(\vec{x}, \vec{a})$.
We define a horn $H(L, c)$:
\begin{equation}
\vec{x} \in H(L, c) \Longleftrightarrow x^{k} = c^{k}(\vec{L}(t),
\vec{a}, t_{1})
\label{t11}
\end{equation}
for some $t, t_{1} \in [0, 1]$ and fix orientation on this surface:
\begin{equation}
\vec{n} || \left( \frac{\partial \vec{c}}{\partial t_{1}}
\times \frac{\partial \vec{c}}
{\partial t} \right)
\label{t1111}
\end{equation}
We are going to show that fluxes ${\cal B}, {\cal E}$ through these
homotopy horns are equal to loop integrals:
\begin{equation}
\int_{H(L, c)} (\vec{B}_{a} \cdot \vec{n}) d \sigma =
\int_{L} f_{a}^{r} d x^{r}
\label{t12}
\end{equation}
\begin{equation}
\int_{H(L^{*}, c^{*})}(\vec{E}_{a} \cdot \vec{n}) d \sigma =
\int_{L} .^{*}f_{a}^{r} dx^{r}
\label{t13}
\end{equation}
with
\begin{equation}
f_{a}^{r}(x) = \int_{c} B_{a}^{k} \varepsilon^{kij}
\frac{\partial y^{j}}{\partial x^{r}} dy^{i}
\label{t14}
\end{equation}
\begin{equation}
^{*}f_{a}^{r}(x) = \int_{c^{*}} E_{a}^{k} \varepsilon^{kij}
\frac{\partial y^{j}}{\partial x^{r}} dy^{i}
\label{t15}
\end{equation}
$L, L^{*}$ and $c, c^{*}$ need not be related. At this stage we need
not specify in what gauge $\vec{B}_{a}, \vec{E}_{a}$ are given.
Eqns.(\ref{t12}), (\ref{t13}) are consequence of a simple observation.
Take any antisymmetric tensor $T_{ij}$ and define
\begin{equation}
g^{r}(x) = \int_{y \in c(\vec{x}, \vec{a})} T^{ij}(y)
\frac{\partial y^{j}}{\partial x^{r}} dy^{i}
\label{t16}
\end{equation}
Then
\begin{equation}
\int_{L} g^{r}(x) dx^{r} = \int dt dt_{1}
\frac{\partial y^{i}}{\partial t_{1}}
\frac{\partial y^{j}}{\partial t}
\varepsilon^{ijk} T^{k}(y)
\label{t17}
\end{equation}
where
\begin{equation}
T^{ij} = \varepsilon^{ijk} T^{k}
\label{t18}
\end{equation}
\begin{equation}
y^{i} = c^{i}(L(t), x_{0}, t_{1})
\label{t19}
\end{equation}
\begin{equation}
\int_{L} g^{r}(x) dx^{r} =
\int_{H(L, c)} (\vec{T} \cdot \vec{n}) d \sigma
\label{t110}
\end{equation}
Replacement $g \longrightarrow f or^{*}f$
and $T^{k} \longrightarrow B^{k} or E^{k}$, gives eqns.(\ref{t12}) and
(\ref{t13}), respectively.\\
If we specify $\vec{B}_{a}$, $\vec{E}_{a}$ to be in
a gauge defined through $c$ from eqn.(\ref{t14}), then $f_{a}^{r}(x)$
is a potential in this gauge. Still there is
a vast choice of homotopies $c^{*}$ defining dual potential
$,^{*}f$. Let us consider Dirac brackets of ${\cal E}, B$
in c-gauge:
\begin{eqnarray}
\left[ {\cal B}_{d}, {\cal E}_{c} \right]_{D} =
\left[ \int_{L} f_{d}^{r} dx^{r}, \int_{L^{*}}.^{*}f_{c}^{s} dy^{s}
\right]_{D} = \nonumber \\
= \left[ \int_{L} f_{d}^{r} dx^{r}, \int_{L^{*}} dy^{s}
\int_{c^{*}(y, a^{*})}dz^{i} \frac{\partial z^{j}}{\partial y^{s}}
\varepsilon^{ijk} E_{c}^{k}(z) \right]_{D}
\label{t111}
\end{eqnarray}
Using (\ref{d17}) one gets, after some algebra, the following
expression:
\begin{eqnarray}
\left[ {\cal E}_{c}^{H^{*}}, {\cal B}_{d}^{H} \right]_{D} =
\delta_{cd} N(L; H^{*}) + \nonumber \\
+ g c_{cdg} \int_{x \in L} dx^{r} f_{g}^{r}(x)
N(c(x,a); H^{*})
\label{t3}
\end{eqnarray}
where
\begin{equation}
N(L; H^{*}) = \sum_{t} sgn \left( \frac{\partial \vec{L}(t_{1})}{\partial t_{1}}
\cdot \vec{n}_{H^{*}}(t_{2}, t_{3}) \right)
\label{t4}
\end{equation}
\begin{eqnarray}
N(c(\vec{x}, a); H^{*}) =
\sum_{\tau(x)} sgn \left( \frac{ \partial \vec{c}(
\vec{x}, \vec{a}, \tau_{1})}{\partial \tau_{1}} \cdot \vec{n}_{H^{*}}(\tau_{2},
\tau_{3}) \right)
\label{t5}
\end{eqnarray}
with $t_{i}$, $\tau_{i}(x)$ being the solutions of the following equations:
\begin{equation}
\vec{c}^{*}(L^{*}(t_{2}), \vec{a}^{*}, t_{3}) = \vec{L}(t_{1})
\label{t6}
\end{equation}
\begin{equation}
\vec{c}^{*}(L^{*}(\tau_{2}), \vec{a}^{*}, \tau_{3}) = \vec{c}(\vec{x}, \vec{a},
\tau_{1})
\label{t7}
\end{equation}
and $\vec{n}_{H^{*}}$ being normal to a horn $ H^{*} \equiv H(L^{*}, c^{*})$
(comp.eqns (\ref{t11}), (\ref{t1111})).\\
The conditions (\ref{t6}) or (\ref{t7}) are fulfilled whenever the surface of
$H^{*}$ is pierced by loop $L$ or homotopy curve $c(\vec{x}, \vec{a})$,
respectively. $N$'s in eqns (\ref{t4}), (\ref{t5}) denote net numbers of
piercings.\\
The abelian part of (\ref{t3}) has been already discussed \cite{l5} for the
radial gauge; for abelian theories it leads to t'Hooft algebra \cite{ll4}. 
The non-abelian part can be
expressed through surface integrals. Call $K_{N}$ part of a loop $L$,
characterized by $N(c; H^{*}) = N$, $N$ fixed ($K_{N}$ can consist of disjoint
pieces). We have $L = \sum_{N} K_{N}$ and corresponding horn surface:
\begin{equation}
H(L, c) = \bigcup_{N} H(K_{N}, c)
\label{t8}
\end{equation}
where
\begin{equation}
x \in H(K_{N}, c) \Longleftrightarrow \vec{x} = \vec{c}(\vec{y}, \vec{a}, t)
\label{t9}
\end{equation}
for some $Y \in K_{N}$ and $t \in [0, 1]$. Evidently eqn(\ref{t12}) holds for
$H(K_{N}, c)$ so that eqn(\ref{t3}) can be rewritten as:
\begin{eqnarray}
[{\cal E}_{c}^{H^{*}}, {\cal B}_{d}^{H}] = \delta_{cd} N(L; H^{*}) +
\nonumber \\
+ g c_{cdg} \sum_{N} N \int_{H(K_{N}, c)} \vec{B}_{g} \cdot \vec{n} d \sigma
\label{t20}
\end{eqnarray}
Let us add, that in fact eqn(\ref{t20}) holds for any surface $S$, not
necessarily a horn $H^{*}(L^{*}, c^{*})$. $H^{*}$ is useful if we want to keep
relation with loop integrals over dual potential (see eqns (\ref{t13}),
(\ref{t15})). More generally, we have:
\begin{eqnarray}
[{\cal E}_{c}^{S}, {\cal B}_{d}^{H}] = \delta_{cd} N(L; S)
+ g c_{cdf} \sum_{N} N \int_{H(K_{N}, c)} \vec{\cal{B}}_{f} \cdot \vec{n}
d \sigma
\label{t21}
\end{eqnarray}
Let us consider now fluxes ${\cal E}_{c}^{(S_{1})}$, ${\cal E}_{c}^{(S_{2})}$.
Surfaces $S_{i}$ are parametrized by given $s_{i}(t_{1}, t_{2})$:
\begin{equation}
x \in S_{i} \Longleftrightarrow \vec{x} = \vec{s}_{i}(t_{1}, t_{2})
\label{t22}
\end{equation}
for some $(t_{1}, t_{2}) \in [0, 1]$.\\

The Dirac bracket of ${\cal E}_{c}^{(S_{1})}$, ${\cal E}_{d}^{(S_{2})}$ -
calculated in $c$-gauge - is given by the following expression:
\begin{eqnarray}
[{\cal E}_{c}^{(S_{1})}, {\cal E}_{d}^{(S_{2})}] = \nonumber \\
g C_{cdf}
\left[ \int_{s_{1} \in S_{1}} \vec{E}_{f}(s_{1}) \vec{n}_{S_{1}}(s_{1})
N(c(\vec{s}_{1}, \vec{a}); S_{2}) d \sigma \right. + \nonumber \\
+ \left. \int_{s_{2} \in S_{2}} \vec{E}_{f}(s_{2}) \vec{n}_{S_{2}}(s_{2})
N(c(\vec{s}_{2}, \vec{a}); S_{1}) d \sigma \right]
\label{t23}
\end{eqnarray}
where $N$'s are the net numbers of piercings:
\begin{eqnarray}
N(c(\vec{s}_{1}, \vec{a}); S_{2})
= \sum_{t_{i}(s_{1})} sgn \left(
\frac{ \partial \vec{c}(\vec{s}_{1}, \vec{a}, t_{3})}{\partial t_{3}} \cdot
\vec{n}_{S_{2}}(t_{1}, t_{2}) \right)
\label{t24}
\end{eqnarray}
with $t_{i}(s_{1})$ being solutions of the following equation:
\begin{equation}
\vec{c}(\vec{s}_{1}, \vec{a}, t_{3}) = \vec{s}_{2}(t_{1}, t_{2})
\label{t25}
\end{equation}
Eqn (\ref{t25}) is fulfilled whenever, for a given $s_{1} \in S_{1}$ the
homotopy curve $c(\vec{s}_{1}, \vec{a})$ crosses the surface $S_{2}$.
Changing $s_{1} \rightarrow s_{2}$, $S_{1} \rightarrow S_{2}$ in eqns
(\ref{t24}), (\ref{t25}) one gets $N$ from the second integral on the r.h.s.
of eqn(\ref{t23}). 
In what follows we shall limit ourselves to surfaces being tangent
at most along a curve to any convolution of homotopy curves  $c(\vec{x},
\vec{a})$ considered in a given gauge. This in turn means that the net
number of piercings given by eqn.(\ref{t24}) is undefined at most along
a curve so that surface integrals in eqn.(\ref{t23}) may have a meaning.
Making in (\ref{t23}) transition $S_{2} \rightarrow S_{1}$
we get for $S_{1} = S_{2} = S$:
\begin{eqnarray}
[{\cal  E}_{c}^{(S)}, {\cal E}_{d}^{(S)}]_{D}
= 2 g c_{cdf} \int_{s \in S}
\vec{E}_{f}(s) \cdot \vec{n}_{S}(s) N(c(s, a); S) d \sigma
\label{t26}
\end{eqnarray}
In this case there is always at least one common point of $c(\vec{s}, \vec{a})$
and $S$, as $c(\vec{s}, \vec{a})$ ends on $s \in S$. The weight of this end-
point contribution to $N$ is $\frac{1}{2}$ as can be seen from the limiting
transition $S_{1} \rightarrow S_{2}$ in eqn(\ref{t23}). Therefore, for any fixed
$s \in S$:
\begin{eqnarray}
2 N(c(s,a); S) = sgn \frac{\partial \vec{c}(\vec{s}, \vec{a}, t_{3})}{\partial
t_{3}} |_{t_{3}=1} \cdot \vec{n}_{S}(\vec{s}) + \nonumber \\
+ 2 \sum_{t; t_{3} \ne 1, s \ne s(t_{1}, t_{2})} sgn \left( \frac{\partial
c(\vec{s}, \vec{a}, t_{3})}{\partial t_{3}} \cdot \vec{n}_{S}(t_{1}, t_{2})
\right)
\label{t27}
\end{eqnarray}
with
\begin{equation}
\vec{c}(\vec{s}, \vec{a}, t_{3}) =  \vec{s}(t_{1}, t_{2})
\label{t28}
\end{equation}
Eqns (\ref{t21}), (\ref{t26}) together with the trivial bracket
\begin{equation}
[{\cal B}_{c}, {\cal B}_{d}]_{D} = 0
\label{t29}
\end{equation}
do not form closed algebra for any chosen $H(L, c)$ and $S$. They can be
however replaced by a set of closed algebras on the properly chosen parts
of $H(L, c)$ and $S$. This will not be discussed here. Let us
conclude with a choice of such $H(L_{0}, c)$ and $S_{0}$ that
\begin{equation}
L_{0} \bigcup S_{0} = \emptyset
\label{t30}
\end{equation}
i.e. abelian part does not contribute to (\ref{t21}). Moreover, put $2 N$
in eqn(\ref{t26}) and $N$ in eqn(\ref{t21}) equal to $1$. (Example: in the
Fock-Schwinger gauge take $H(L_{0}, C)$ to be a cone and $S_{0}$ to be any
planar surface containing elliptic section of $H_{0}$). In such a case we have:
\begin{equation}
[{\cal E}_{a}^{S_{0}}, {\cal B}_{b}^{H_{0}}]_{D} = g c_{abc} {\cal B}_{c}^{H_{0}}
\label{t31}
\end{equation}
\begin{equation}
[{\cal E}_{a}^{S_{0}}, {\cal E}_{b}^{S_{0}}]_{D} = g c_{abc} {\cal E}_{c}^{S_{0}}
\label{t32}
\end{equation}
\begin{equation}
[{\cal B}_{a}^{H_{0}}, {\cal B}_{b}^{H_{0}}]_{D} = 0
\label{t33}
\end{equation}
If we took $S^{0}$ to be closed surface surrounding $\vec{a}$, then (\ref{t32})
still holds and
is the algebra of charges  contained in its interior, $V_{0}$. The 
question whether the algebra holds for any closed surface will be 
discussed in the next chapter.

\section{Local charge algebras}
\setcounter{equation}{0}
It seems interesting to investigate whether colour electric flux algebra
from (\ref{t32}) can be extended to the general case of two arbitrary
closed surfaces; if it were so, then we might be able to establish
local colour charge algebras smilar to well known flavour current 
algebras of the traditional quark model. \\
Let us take closed surfaces $S_{i}\in V_{-}$ ($V_{-}=V - P(\vec{a})$,
comp. eqn.(\ref{d10})) which are boundaries of 3-dimensional, open regions
$V_{i} \in V$. Next, denoting an outward flux of $\vec{E}_{a}$ through 
$S_{i}$ as ${\cal E}_{a}(V_{i})$, we shall check commutation relations 
of colour electric fluxes
\begin{equation}
{\cal E}_{a}(V_{i}) =_{def} {\cal E}_{a}^{S_{i}}
\label{dzia1}
\end{equation}
for two situations: a)$V_{1}=V_{2}$, ~b) $V_{1} \cap V_{2} = \emptyset$. \\
For the case a) we get from eqns.(\ref{t26}-\ref{t28}) 
\begin{center}
a) ~~$V_{1} = V_{2}$
\end{center}
\begin{eqnarray}
if ~~P(\vec{a}) \in V_{1}, ~~[{\cal E}_{a}(V_{1}), {\cal E}_{b}(V_{1})]_{D} 
= g c_{abc} {\cal E}_{c}(V_{1}) 
\label{dzia2} \\
if ~~P(\vec{a}) \not\in V_{1}, ~~[{\cal E}_{a}(V_{1}), {\cal E}_{b}(V_{1})]_{D} = - 
g c_{abc} {\cal E}_{c}(V_{1})
\label{dzia3}
\end{eqnarray}
For the case b) we get from eqns.(\ref{t23}-\ref{t25})
\begin{center}
b) ~~$V_{1} \cap V_{2} = \emptyset$
\end{center}
\begin{eqnarray}
if~~ P(\vec{a}) \in V_{1}, P(\vec{a}) \not\in V_{2}, ~~[{\cal E}_{a}(V_{1}), 
{\cal E}_{b}(V_{2})]_{D} 
= g c_{abc} {\cal E}_{c}(V_{2})
\label{dzia4} \\
if ~~P(\vec{a}) \not\in V_{1} V_{2}, ~~[{\cal E}_{a}(V_{1}), {\cal E}_{b}(V_{2})]_{D} = 0
\label{dzia5}
\end{eqnarray}
If we assumed that $\vec{E}$ is regular at $\vec{r}=\vec{a}$, then eqn.(\ref{dzia4})
would mean that the electric charges defined by (\ref{dzia1})
do not form local algebra. 
However, such an assumption is not sound; eqns.(\ref{dzia2}-\ref{dzia5}) exhibit 
essential role of the point $P(\vec{a})$ despite the fact that $P(\vec{a})\not\in
S_{i}$ in our derivations. Having this in mind let us consider a sequence 
$\{V_{0}^{k}\}$ of open balls containing $\vec{a}$. Let their boundaries 
$S_{0}^{k}$ have radii $\varepsilon_{k} \longrightarrow 0$ for $k \longrightarrow
\infty$ (please notice that $\varepsilon_{k} = 0$ is never reached as $S_{0}^{k}
\subset V_{-})$. Let us define
\begin{equation}
e_{c}^{k} = \int_{S_{0}^{k}} \vec{E}_{c} \vec{n} d \sigma
\label{dzia6}
\end{equation}
and assume existence of limit
\begin{equation}
e_{c} = \lim_{k \longrightarrow \infty} e_{c}^{k}
\label{dzia7}
\end{equation}
For any fixed surface $S$ eqns.(\ref{t23}-\ref{t25}) lead to the following relation
\begin{equation}
[e_{a}, {\cal E}_{b}^{S}]_{D} =_{def} \lim_{k \longrightarrow \infty}
[e_{a}^{k}, {\cal E}_{b}^{S}]_{D} = g c_{abc} {\cal E}_{b}^{S}
\label{dzia8}
\end{equation}
and, as a consequence,
\begin{equation}
[e_{a}, {\cal E}_{b}(V_{i})]_{D} = g c_{abc} {\cal E}_{c}(V_{i})
\label{dzia9}
\end{equation}
Moreover,
\begin{equation}
[e_{a}, e_{b}]_{D} =_{def} \lim_{k, r \longrightarrow \infty} [e_{a}^{k}, 
e_{b}^{r}]_{D} = g c_{abc} e_{c}
\label{dzia10}
\end{equation}
and, due to eqn.(\ref{d17}),(\ref{d19})
\begin{equation}
[e_{a},A_{b}^{i}(\vec{x})]_{D} = g c_{abc} A_{c}^{i}(\vec{x})
\label{dzia10a}
\end{equation}
\begin{equation}
[e_{a},E_{b}^{i}(\vec{x})]_{D} = g c_{abc} E_{c}^{i}(\vec{x})
\label{dzia10b}
\end{equation}
for any $\vec{x} \in V_{-}$.\\
Now we are in a position to introduce charges associated with the new field $T_{a}$
defined below
\begin{eqnarray}
Q_{a}^{T}(V_{1}) = \int_{S_{1}} \vec{T}_{a} \cdot \vec{n} d \sigma
\label{dzia11}\\
\vec{T}_{a}(\vec{r}) = \vec{E}_{a}(\vec{r}) - \frac{1}{4 \pi} \frac{\vec{r} -
\vec{a}}{|\vec{r} - \vec{a}|^{3}} e_{a}, ~~\vec{r} \in V_{-}
\label{dzia12}
\end{eqnarray}
so that
\begin{equation}
Q_{a}^{T}(V_{1}) = {\cal E}_{a}(V_{i}) - g(V_{1}) e_{a}
\label{dzia14}
\end{equation}
where
\begin{eqnarray}
g(V_{1}) = 1 ~~ if ~\vec{a} \in V_{1}
\nonumber \\
g(V_{1}) = 0 ~~ if ~\vec{a}\not\in V_{1}
\label{dzia15}
\end{eqnarray}
Using eqns.(\ref{dzia1}-\ref{dzia15}) one can check that $Q_{a}^{T}(V_{i})$
satisfy local algebra
\begin{equation}
[Q_{a}^{T}(V_{1}), Q_{b}^{T}(V_{1})]_{D} = -g c_{abc} Q_{c}^{T}(V_{1}) 
\label{dzia16}
\end{equation}
\begin{center}
for any fixed open $V_{1} \subset V$ \\
\end{center}
and
\begin{equation}
[Q_{a}^{T}(V_{1}), Q_{b}^{T}(V_{2})]_{D} = 0
\label{dzia17}
\end{equation}
\begin{center}
for $V_{1} \cap V_{2} = 0$
\end{center}
Eqns.(\ref{dzia2}), (\ref{dzia3}) together with eqns.(\ref{dzia12}-\ref{dzia17}) clearly
demonstrate existence of a point term at $\vec{r}=\vec{a}$. This leads to non-integrable singularity
in Hamiltonian, unless $e_{a}=0$. In quantum theory it would mean that we choose such a subspace
of states, ${\cal H}$, that
\begin{equation}
e_{a}|\Psi> = 0 
\label{dzia18}
\end{equation}
for any $|\Psi> \in {\cal H}$ and for any "$a$".\\
Then eqns.(\ref{dzia10a}),(\ref{dzia10b}) lead to 
\begin{equation}
E_{a}^{i}(\vec{x}) |\Psi> = A_{a}^{i}(\vec{x}) |\Psi> = 0
\label{dzia19}
\end{equation}
for any $\vec{x} \in V_{-}$ and any $|\Psi> \in {\cal H}$, i.e. theory is trivial unless
$e_{a} \ne 0$. Therefore in order to avoid obstruction  to the Gauss law we have to exclude a curve
from the region $V$ and to deal with the region $V_{-c_{0}}$
\begin{equation}
V_{-c_{0}} = V - \{ c(\vec{R}_{0} \in S, \vec{a}) \}
\label{dzia20}
\end{equation}
where $c_{0} = c(\vec{R}_{0} \in S, \vec{a})$ is any curve joining $\vec{a}$ with some chosen point 
$\vec{R}_{0}$ at the boundary $S$ of the region $V$. The curve $c_{0}$ should belong to a family of 
curves used for gauge fixing so that nonabelian parts of Dirac brackets (eqns.(\ref{d17}),(\ref{d19}))
would not mix dynamical variables on $c_{0}$ with those on $V_{-c_{0}}$. The choice of $V_{-c_{0}}$
makes the use of $e_{a}$ unnecessary as closed surfaces around $\vec{a}$ cannot be contained in
$V_{-c_{0}}$. 
Now, whether curve $c_{0}$ will be a regular or singular line of field $E$ can be in principle derived 
from dynamics of the theory in the region $V_{-c_{0}}$.

\begin{center}
{\bf Acknowledgements}
\end{center}
We gratefully acknowledge the contributions of Professor A.A.Anselm to the early
stages of this work.
We are indebted to Professors Gregory Korchemsky, Wojciech Kr\'olikowski and Martin Reuter
for interesting remarks and discussions.
Two of us (A.J., L.{\L}.) would like to thank the Royal Society for generously
supporting our stay at Birkbeck College, London, where this work was begun.
\appendix
\section{Appendix}
\setcounter{equation}{0}
We shall show that $A = f_{\rho}^{c}$ given by eqn.(\ref{t15b})
of theorem 2 satisfies eqn.(\ref{at1}) if eqn.(\ref{at3}) is satisfied.
\\
Let us start with formula
\begin{equation}
F_{\alpha \rho} = f_{\rho, \alpha}^{c} - f_{\alpha, \rho}^{c} =
G_{\alpha \rho} - \int \frac{\partial c^{\mu}}{\partial x^{\rho}}
\frac{\partial c^{\sigma}}{\partial x^{\alpha}}
\frac{\partial c^{\nu}}{\partial t} \cdot [G_{\mu \nu, \sigma} +
G_{\nu \sigma, \mu} + G_{\sigma \mu, \nu}] dt
\label{ap1}
\end{equation}
which is obtained from (\ref{t15b}) in a manner similar
to that applied in the proof of theorem 1.\\
Next, using (\ref{at3}) we get from (\ref{ap1}):
\begin{equation}
G_{\alpha \rho} - F_{\alpha \rho} = - i g \left[ [f_{\sigma}^{c},
G_{\mu \nu}] + [f_{\mu}^{c}, G_{\nu \sigma}] \right]
\label{ap2}
\end{equation}
(The term $[f_{\nu}^{c}, G_{\sigma \mu}]$ does not contribute, because
$\frac{\partial c}{\partial t} \cdot f^{c} = 0$).\\
Substituting $f^{c}$ from eqn.(\ref{t15b}) we get
\begin{equation}
G_{\alpha \rho} - F_{\alpha \rho} = - i g (I_{\rho \alpha} -
I_{\alpha \rho})
\label{ap3}
\end{equation}
where
\begin{equation}
I_{\rho \alpha} = \int \frac{\partial c_{t}^{\mu}}{\partial x^{\rho}}
\frac{\partial c_{t}^{\sigma}}{\partial x^{\alpha}}
\frac{\partial c_{t}^{\nu}}
{\partial t} \frac{\partial c^{\delta}(c_{t}, x_{0}, \tau)}
{\partial c_{t}^{\sigma}} \frac{\partial c^{\beta}(c_{t}, x_{0}, \tau)}
{\partial \tau} d t d \tau [G_{\beta \delta}(c(c_{t}, x_{0}, \tau)),
G_{\mu \nu}(c_{t})]
\label{ap4}
\end{equation}
and $c_{t} = c(x, x_{0}, t)$\\
Applying eqn.(\ref{at15}) we get
\begin{equation}
I_{\rho \alpha} = \int_{0}^{1} d t \int_{0}^{t} d h \frac{\partial
c_{t}^{\mu}}{\partial x^{\rho}} \frac{\partial c_{t}^{\nu}}
{\partial t} \frac{\partial c_{h}^{\delta}}{\partial x^{\alpha}}
\frac{\partial c_{h}^{\beta}}{\partial h} [G_{\beta \delta}(c_{h}),
G_{\mu \nu}(c_{t})]
\label{ap5}
\end{equation}
and, after a suitable change of variables,
\begin{equation}
I_{\alpha \rho} = - \int_{0}^{1} d h \int_{0}^{h} d t
\frac{\partial c_{t}^{\mu}}{\partial x^{\rho}}
\frac{\partial c_{t}^{\nu}}{\partial t}
\frac{\partial c_{h}^{\delta}}{\partial x^{\alpha}}
\frac{\partial c_{h}^{\beta}}{\partial h}
[G_{\beta \delta}(c_{h}), G_{\mu \nu}(c_{t})]
\label{ap6}
\end{equation}
so that
\begin{equation}
I_{\rho \alpha} - I_{\alpha \rho} = [f_{\rho}^{c}, f_{\alpha}^{c}]
\label{ap7}
\end{equation}
Then from (\ref{ap3})
\begin{equation}
G_{\alpha \rho} = F_{\alpha \rho} + i g [f_{\alpha}^{c}, f_{\rho}^{c}]
\label{ap8}
\end{equation}
i.e. eqn.(\ref{at1}) as desired. Eqn.(\ref{at5}) trivially follows
from (\ref{t15a}); it is enough to repleace $T$ by $G$ in (\ref{t15b})
and use the result (\ref{t151}) of the lemma.


\begin{thebibliography}{30}
\bibitem{l1}V.A.Fock, Sov.J.Phys.\underline{12}, 404 (1937)
\bibitem{l2}J.Schwinger, Phys.Rev. \underline{82}, 664 (1951)
\bibitem{l3}C.Cronstrom, Phys.Lett. \underline{90B}, 267(1980) \\
V.Novikov, M.Shifman, A.Vainshtein, V.Zakharov, Fortschr. der Physik,
\underline{32}, 585 (1984)
\bibitem{ll1} S.V.Ivanov, G.P.Korchemsky, Phys.Lett. \underline{B154}
(1985)197\\
S.V.Ivanov et al  Sov. J. Nucl. Phys. 44(1986)230\\
G.P.Korchemsky, A.V.Radyushkin, Phys. Lett. \underline{B171}(1986)459\\
S.V.Ivanov, Fiz.Elem.Chastits At.Yadra 21(1990)75
\bibitem{NX} P.Gaete, Z.Phys.C76(1997)335\\
L.Prokhorov, S.Shabanov, Int.J.Mod.Phys.A7(1992)7815
\bibitem{l4}M.B.Halpern, Phys.Rev. \underline{19D}, 517 (1979)
\bibitem{l5}A.Hosoya, K.Shigemoto, Progr.of Theor.Phys. \underline{65}
, 2008(1981)
\bibitem{l6}S.Weinberg, Gravitation and Cosmology, ch.4.11, ed.John
Wiley and Sons, New York
\bibitem{l7}K.Maurin, Analysis, part II, ch.XIV.3, ed.PWN Warszawa
and D.Reidel, Dordrecht
\bibitem{l8}R.Gambini, A.Trias, Phys.Rev. \underline{D23}, 553(1981)
\bibitem{l9}S.Weinberg, The Quantum Theory of Fields, Cambr.Univ.Press,
(1995) \\
A.J.Hanson, T.Regge, C.Teitelboim, Academia Nazionale dei Lincei, Rome
1974
\bibitem{ll4} M.B.Halpern, Phys.Lett. \underline{B81}, 245(79)\\
G.t'Hooft, Nucl.Phys. \underline{B153}, 141(1979)\\
A.Caticha, Phys.Rev. \underline{D37}, 2323(1988)\\
L.Leal, Mod.Phys.Lett. \underline{A11}, 1107(1996)


\end{thebibliography}
\end{document}